\documentstyle[11pt,iau177,twoside,epsf]{article}

\begin{document}
\title{The Effelsberg Search for Pulsars in the Galactic Centre}
 \author{M.~Kramer, B.~Klein, D.~Lorimer, P.~M\"uller, A.~Jessner,
R.~Wielebinski}
\affil{Max-Planck-Institut f\"ur Radioastronomie, Bonn, Germany}

\begin{abstract}
We report the status of a search for pulsars in the Galactic Centre,
using a completely revised and improved high-sensitivity double-horn
system at 4.85-GHz. We also present calculations about the success
rate of periodicity searches for such a survey, showing that in
contrast to conclusions in recent literature pulsars can be indeed
detected at the chosen search frequency.
\end{abstract}

\noindent
 The detection of radio pulsars in the near vicinity of the Galactic
Centre is apparently hampered by the largely increased scattering of
pulsar signals at the inhomogeneities of the interstellar medium.
This effect cannot be removed by instrumental means but can only be
reduced by observations at a high radio frequencies. The steep spectra
of pulsars require a compromise in the choice of the search frequency
to be used. As a consequence, we had started to search the Galactic
Centre at a frequency of 4.85-GHz using the Effelsberg 100-m radiotelescope
(Kramer et al.~1996, in Proc.~of IAU Colloq.~160, \pasp, p.~13). 

Since recently, we have employed both horns of the high-sensitive double-horn
6cm-frontend ($T_{sys}\approx 25$ K) connected to four 8-channel
filterbanks with $B=8\times80=480$ MHz bandwidth.  The data are
recorded by a new ``48+''-channel backend working under VxWorks with a
maximum datarate of $\sim30$ MB/s (10 MB/s sustained).  In combination
with new reduction software this results in a greatly improved
sensitivity of less than 0.1 mJy for a 12 min integration (Fig.~1).

Recently, Cordes \& Lazio (1997, \apj, 475, 557, hereafter CL97) presented
calculations which indicated that periodicity searches in the Galactic
Centre below 10 GHz would hardly be successful, since scattering will
reduce the pulsed fraction of the pulsar signal at lower frequencies.
Instead, they favoured a complementary imaging approach. It has to be
noted, however, that their calculations made use of a simplified
expression to describe the effect of scattering.  A correct treatment
shows a still significant contribution of power in the higher
harmonics. As a result, the decrease of pulsed emission towards lower
frequencies is much slower than derived by CL97. In fact, we 
demonstrate that pulsars can still be discovered in periodicity
searches at 5 GHz (Fig.~2).

\begin{figure}
\plotfiddle{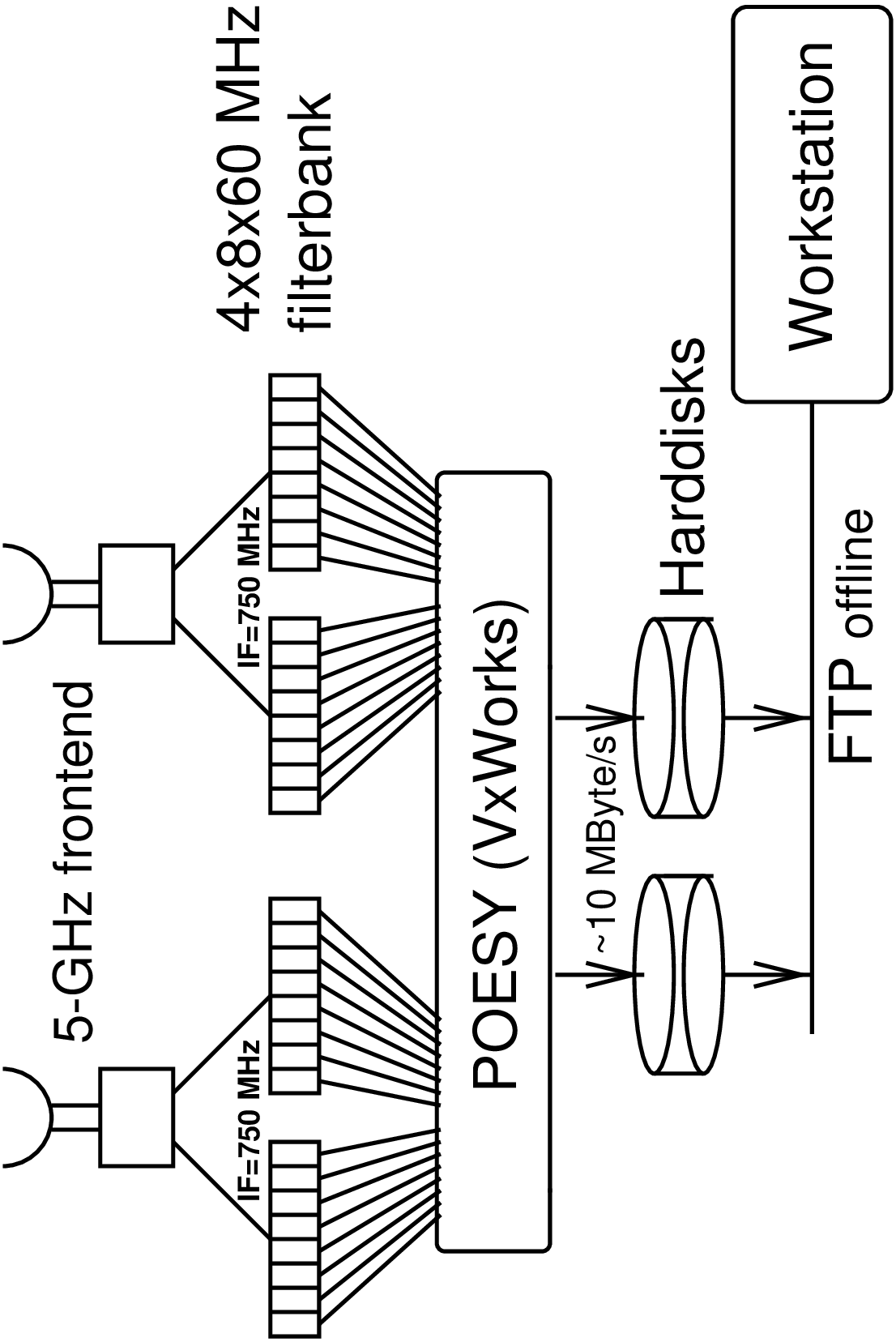}{3.1cm}{270}{30}{40}{-240}{180}
\plotfiddle{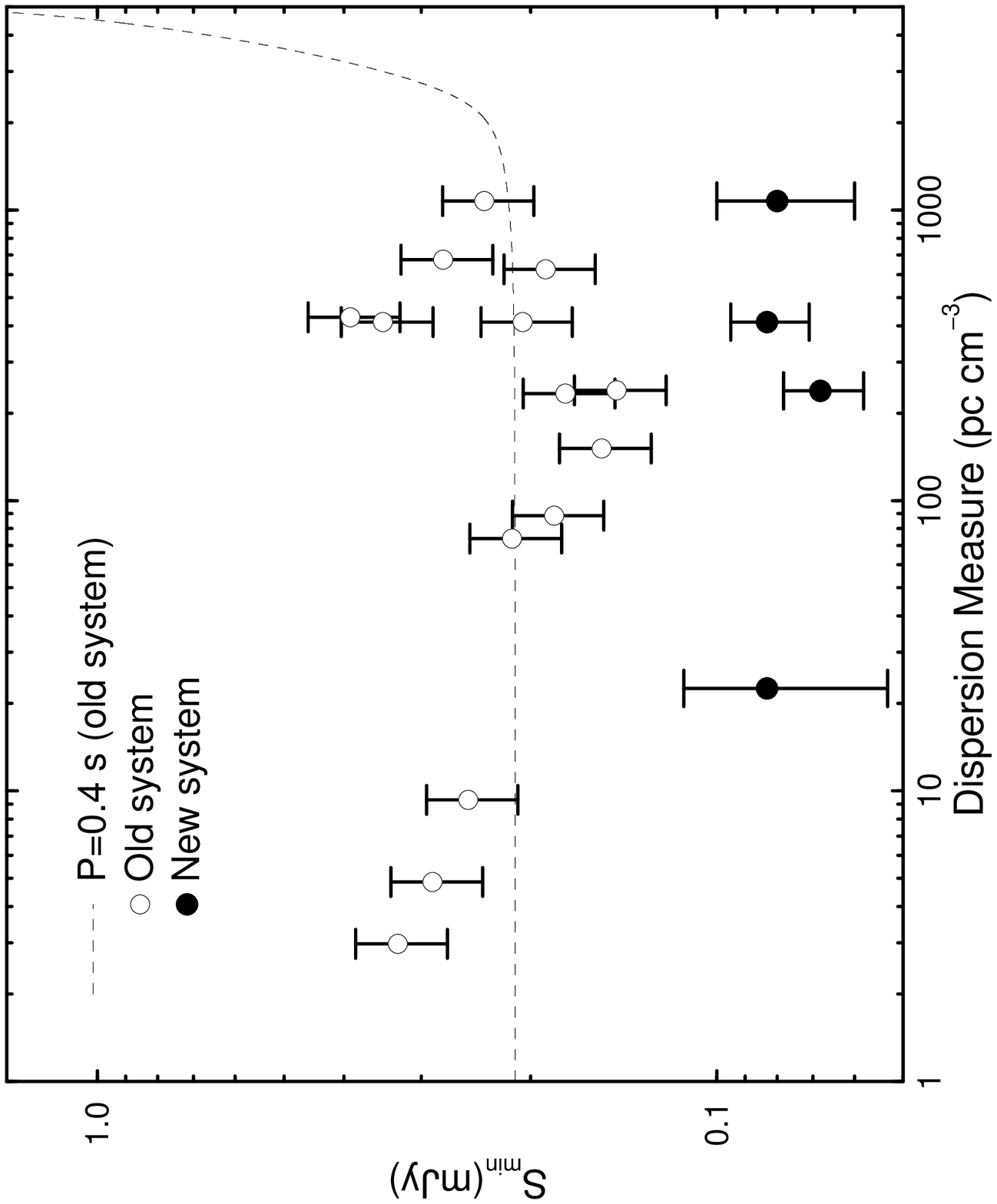}{3.1cm}{270}{35}{45}{0}{250}
\caption{{\it left)} New search system for the 6cm Galactic
Centre survey, {\it right)} improved sensitivity of the new system 
in comparison to that described by Kramer et al.~(1996).}
\end{figure}

\begin{figure}
\plotfiddle{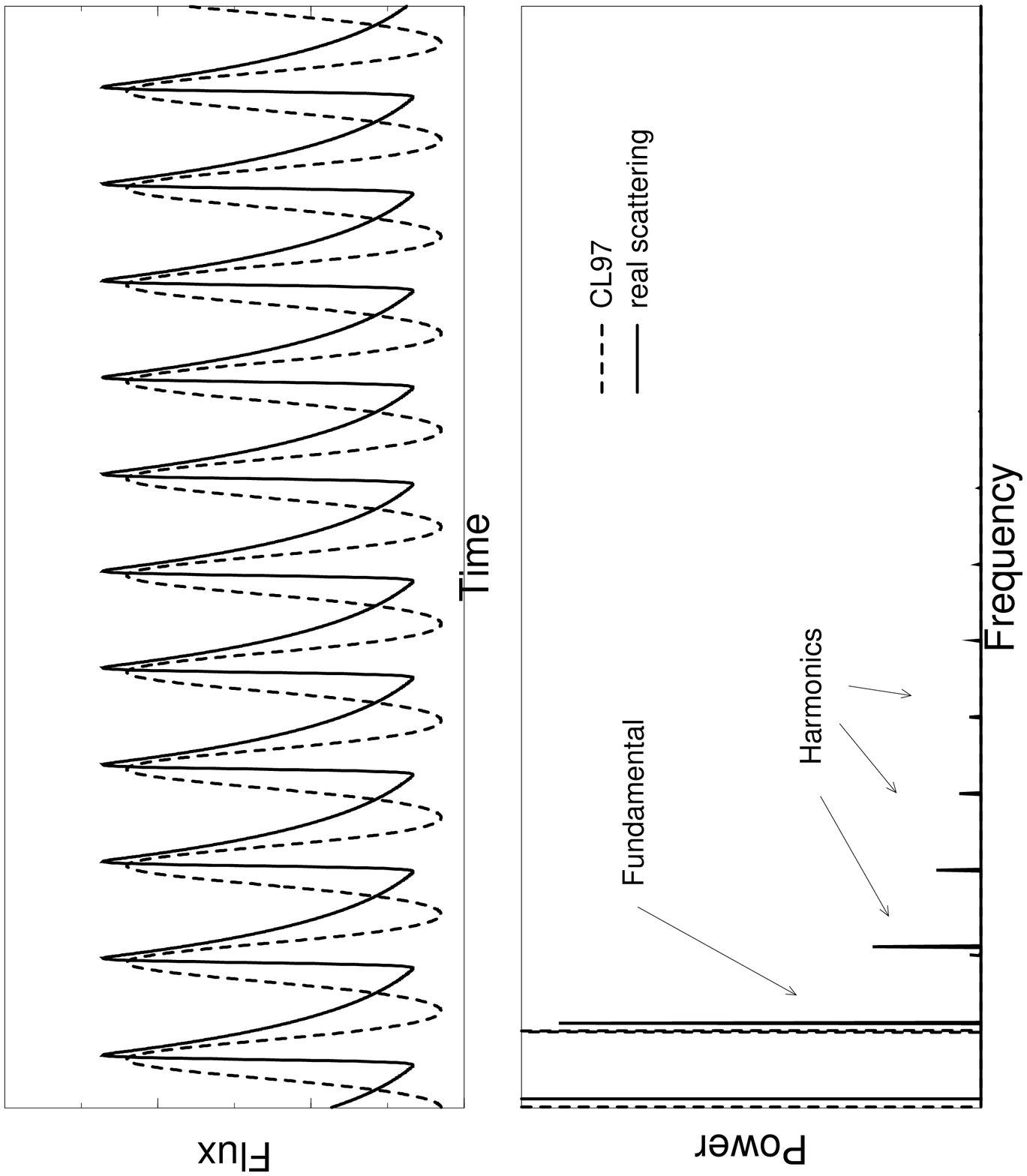}{2.5cm}{270}{30}{35}{-200}{130}
\plotfiddle{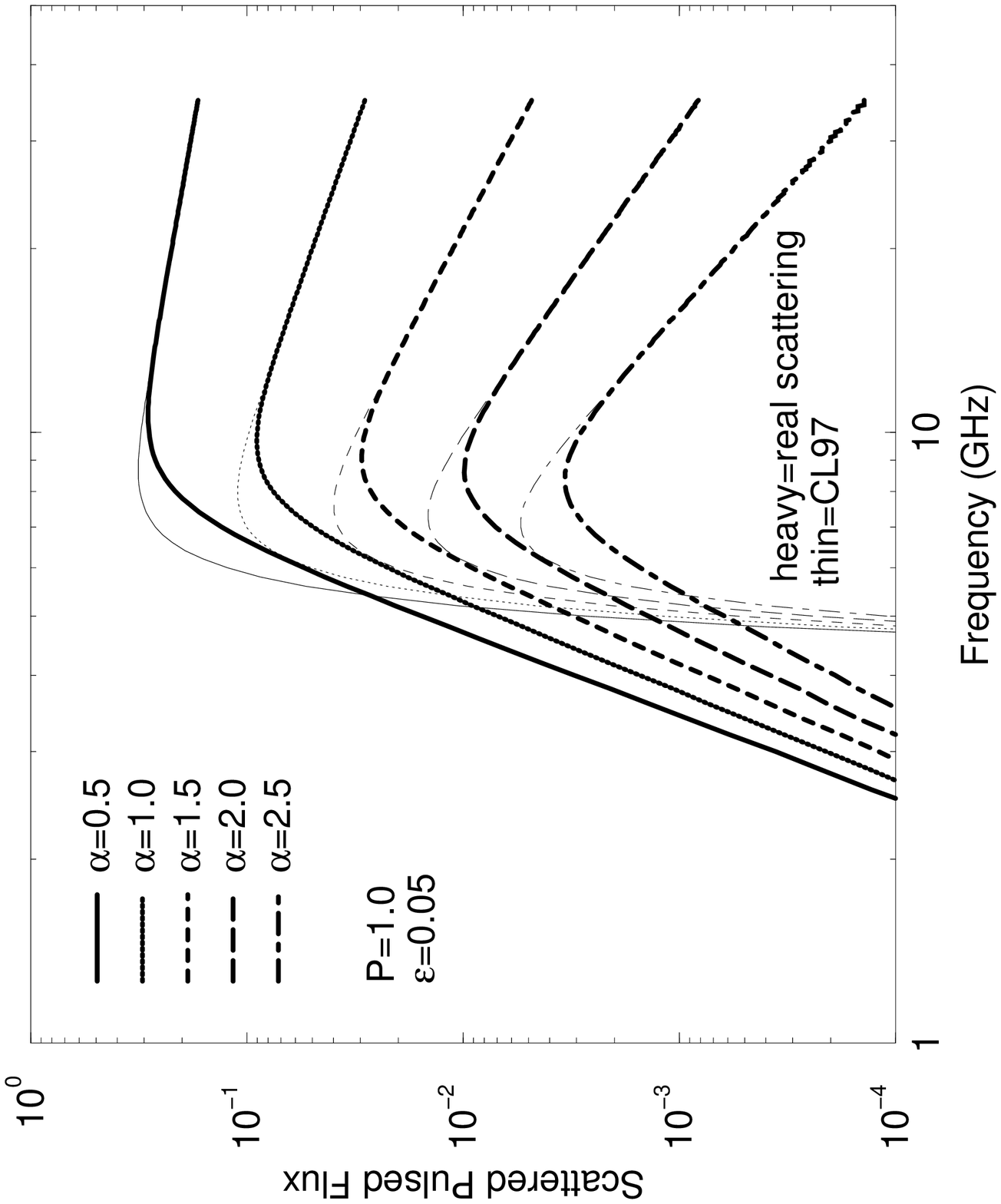}{2.5cm}{270}{35}{45}{-10}{253}
\caption{{\it left)} Effects of pulse scattering on time series and
power spectrum when modeled as by CL97 and when applying the
actual formulae, {\it right)} scattered pulsed flux for a model pulsar
of $P=1$ s and a duty cycle of 5\% for different spectral indices.
In contrast to results by CL97 (thin lines), real scattering (heavy
lines) still produces detectable pulsed fraction at frequencies around
5 GHz.}
\end{figure}

\acknowledgements We thank the receiver group of the MPIfR for
building the superb 6cm-receiver and the corresponding filterbanks.
The digital-group, in particular Thomas Kugelmeier, made significant
contributions in the development of the new backend POESY.


\end{document}